\documentclass[prl,aps,showpacs,twocolumn]{revtex4}
\usepackage{amsmath}
\usepackage{graphicx}

\begin{document}

\title{Quartets and the Current-Phase Structure of a Double Quantum Dot Superconducting Bijunction At Equilibrium}
\date{\today}

\author{D. Feinberg$^{1,2}$, T. Jonckheere$^{3}$, J. Rech$^{3}$,  T. Martin$^{3}$,
  B. Dou\c{c}ot$^4$, R. M\'elin$^{1,2}$ } 

\affiliation{$^1$ Centre National de la Recherche Scientifique, Institut NEEL, F-38042 Grenoble Cedex 9, France}
\affiliation{$^2$ Universit\'e Grenoble-Alpes, Institut NEEL, F-38042 Grenoble Cedex 9, France}
\affiliation{$^3$ Aix-Marseille Universit\'e, Universit\'e de Toulon, CNRS, CPT, UMR 7332, 13288 Marseille, France}
\affiliation{$^4$ Laboratoire de Physique Th\'eorique et des Hautes Energies,
  CNRS UMR 7589, Universit\'es Paris 6 et 7, 4 Place Jussieu, 75252 Paris
  Cedex 05}

\begin{abstract}
The equilibrium current-phase structure of a tri-terminal superconducting Josephson junction (bijunction) is analyzed as a function of the two relevant phases. The bijunction is made of two noninteracting quantum dots, each one carrying a single level. Nonlocal processes coupling the three terminals are described in terms of quartet tunneling and pair cotunneling. These couplings are due to nonlocal Andreev and cotunneling processes through the central superconductor $S_0$, as well as direct interdot coupling. In some cases, two degenerate midgap Andreev states appear, symmetric with respect to the ($\pi,\pi$) point. The lifting of this degeneracy by interdot couplings induces a strong non-local inductance at low enough temperatures. This effect is compared to the mutual inductance of a two-loop circuit. 
\end{abstract}


\maketitle

\section {I. Introduction}
Josephson junctions couple two superconductors by an insulator or normal metal bridge $N$ \cite{Tinkham}. In the latter case, the Josephson effect in a two-terminal $SNS$ junction relies on the coherence of the Andreev reflections at each $NS$ interface, which results at  equilibrium in the Andreev bound states (ABS). Two Andreev reflections, one at each interface, allow one Cooper pair to cross the $SNS$ junction. The ABS dispersion with the phase difference at the junction essentially controls the current-phase (CPR) relationship of the junction. The CPR can be experimentally probed by SQUID interferometry\cite{cpr}, and, more recently, the ABS structure has been directly investigated by microwave spectroscopy\cite{Bretheau,BouchiatABS}. Dot and double-dot set-ups can also be investigated by resonant coupling to a microwave cavity \cite{cottet}.

The present work focuses on the ABS structure at equilibrium of a tri-terminal Josephson \cite{Cuevas,Houzet,Chtchelkatchev,Freyn,Zaikin2012,Zaikin2013,Alidoust,Jonckheere,Pfeffer}. It elucidates its current-phase relation as a function of the two phase variables, hence the name "bijunction". It clarifies the nature of several nonlocal processes occurring in such a structure. This current-phase relation could be probed by methods inspired by those used in the framework of two-terminal junctions. For instance, a two-loop biSQUID geometry has been recently proposed by us \cite{bisquid}. On the other hand, for transparent enough contacts, the Andreev bound states formed within the bijunction could be probed by spectroscopy tools {\cite{Bretheau,BouchiatABS}, or, as recently suggested, using a closeby $NS$ junction\cite{gosselin}.

More specifically, we consider here the case of a bijunction (Figure \ref{DoubleDotBJ}) where each arm is formed by a single level quantum dot\cite{Jonckheere}, made for instance from a single carbon nanotube or nanowire. 
This structure is closely related to hybrid bijunctions made of two quantum dots and normal (instead of superconducting) reservoirs $N_{a,b}$, which have been fabricated either with carbon nanotubes or with semiconducting nanowires, in a $(N_aD_aS_0D_bN_b)$ structure\cite{Nanosquid,hofstetter2009,herrmann2010,das2012}. Indeed, nonlocal processes in double $(N_aS_0N_b)$ hybrid structures connecting one superconductor $S_0$ to two normal metals $N_{a,b}$ have been predicted \cite {byers1995,martin1996,anantram1996,deutscher2000,lesovik2001,recher2001,bouchiat2003,samuelsson2003,melin2004} and explored in experiments\cite {beckmann2004,russo2005,zimansky2006,hofstetter2009,herrmann2010,das2012}, with the prospect of producing entangled pairs of electrons. In the language of quasiparticle scattering, either an electron (hole) impinging on $S_0$ from $N_a$ is normally transmitted as an electron (hole) towards $S_b$, or it is Andreev-transmitted as a hole (electron). The first channel corresponds to tunneling of a quasiparticle through the superconducting gap (so-called "elastic cotunneling" EC), while the second one involves the creation (annihilation) of a Cooper pair in $S_0$ and is a nonlocal (crossed) Andreev process (CAR). 
The latter amounts to split Cooper pairs into entangled singlets\cite{lesovik2001,recher2001,bouchiat2003,samuelsson2003,burset2011}, and is responsible for nonlocal and spin-dependent conductance, while the proof of spin entanglement remains elusive.
The experimental results clearly show the existence of nonlocal processes leading to splitting Cooper pairs from $S_0$ into pairs of quasiparticles in $N_a$, $N_b$.  

In an all-superconducting bijunction, CAR and EC result in new coherent multipair transport channels, that must occur between the three terminals \cite{Cuevas,Houzet,Chtchelkatchev,Freyn,Jonckheere}.
At equilibrium, in a bijunction, the combination of crossed Andreev process at $S_0$ and local Andreev reflection at $S_{a,b}$ builds ABS, which depend on {\it two} phase variables, say $\varphi_a-\varphi_0$, $\varphi_b-\varphi_0$. 
Those states can in particular mediate the simultaneous passage of {\it two} Cooper pairs from $S_0$ towards $S_a$, $S_b$, achieving so-called quartet transport. 

\begin{figure}[tb]
\centerline{\includegraphics[width=1.1\columnwidth]{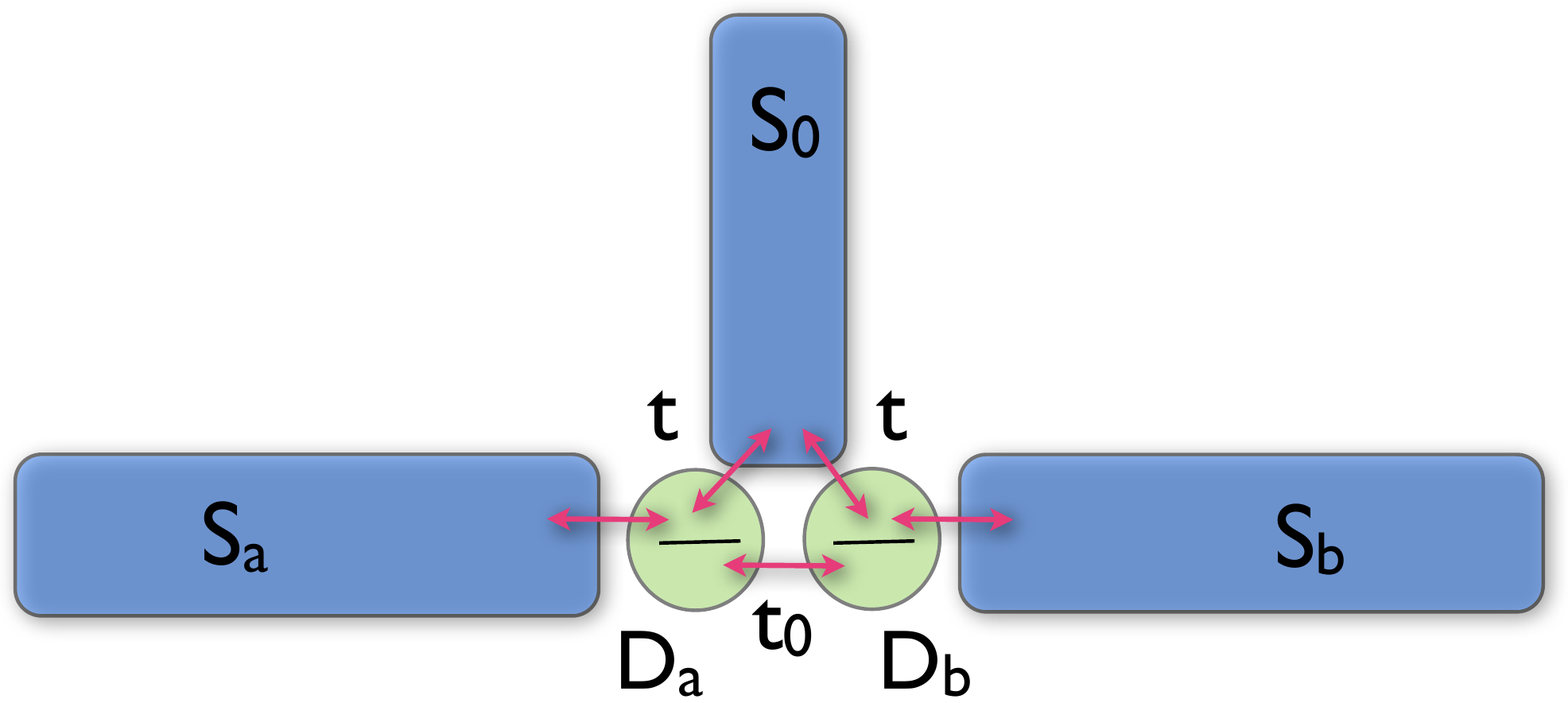}}
\caption{Bijunction considered in this paper, made of two quantum dots $D_a$, $D_b$.  The interdot ($t_{dd}$) and superconductor-dot ($t_{sd}$) hopping parameters are indicated.
\label{DoubleDotBJ}
}
\end{figure}

In the present case of an all-superconducting tri-terminal set-up, these new processes introduce a microscopic coupling between the two junctions \cite{Freyn}. At equilibrium, the general picture is that of Andreev bound states coherently formed on both junctions $a,b$ simultaneously. As a result, the total energy of the bijunction is a $2\pi$-periodic function $E_{BJ}=E(\varphi_{0a}, \varphi_{0b})$ of the phase differences $\varphi_{0a}=\varphi_a-\varphi_0$ and $\varphi_{0b}=\varphi_b-\varphi_0$, and the currents
\begin{equation} 
I_a=\frac{2e}{\hbar}\frac{\partial E_{BJ}}{\partial \varphi_{0a}}\;,\;\;\;
 I_b=\frac{2e}{\hbar}\frac{\partial E_{BJ}}{\partial \varphi_{0b}}
\end{equation}
are both functions of $\varphi_{0a}$ {\it and} of $\varphi_{0b}$. 
In this work we derive the exact current-phase relationship (CPR) in a two-dot bijunction. Due to the tri-terminal geometry, nontrivial midgap states may appear, symmetric with respect to the central ($\pi,\pi$) point. The importance of such states has been recently underlined in Ref. \onlinecite{Akhmerov2014}. We show how the underlying degeneracy is lifted by interdot couplings, directly or through the central supetconductor. This understanding of the CPR should clarify the nature of the nonequilibrium transport, which offers new coherent dc channels in presence of applied voltages, provided the latter are commensurate\cite{Cuevas,Freyn,Jonckheere}, and also nonlocal multiple Andreev incoherent channels\cite{Houzet,Chtchelkatchev}. Subgap anomalies in a diffusive Al-Cu bijunction have indeed been recently observed and interpreted in terms of quartets (see Figure \ref{Bijunctions}b)\cite{Pfeffer}. Notice that a related set-up has been proposed in the context of Majorana fermion physics\cite{vonoppen}. 

Section II defines the model and the exact solution for the ABS, that becomes analytic in the low energy limit. Section III discusses the structure of the ABS states of the bijunction, first in the analytic limit. Section IV provides a discussion of the currents and the resulting nonlocal inductance in the general case, and also considers the role of the circuit inductances when the phases are imposed by a two-loop set-up. 

\section{II. Bijunction with two quantum dots: the model}

Each junction $S_{a(b)}-S_0$ is formed by a quantum dot $D_{a(b)}$ with a single noninteracting level, with energies $E_{a(b)}$ respectively, and a direct coupling between the single levels in $D_{a(b)}$  in the electron-electron channel (Figure \ref{DoubleDotBJ}). Such a coupling  is a simplified way to modelize the connectivity of the nanotube\cite{herrmann2010,burset2011}. The Hamiltonian of the system is written in the Nambu notation $H=H_{S}+H_{DD}+H_{T}$, and performing a gauge transformation to incorporate the superconducting phases $\varphi_j$ in the tunneling term $H_T$ :

\begin{equation}\label{Hamilt}
H_S=\sum_{j=a,b,0}\sum_{k}
\Psi_{jk}^{\dagger}(\xi_{k}\sigma_{z}+\Delta_j \sigma_{x})
\Psi_{jk}, \Psi_{jk}=\left(\begin{array}{c} \psi_{jk,
\uparrow}\\
\psi_{j(-k),\downarrow}^{\dagger}
\end{array} \right)
\end{equation}

\begin{equation}
H_{DD}=\sum_{\alpha=a,b}E_{\alpha}d^{\dagger}_{\alpha} \sigma_{z}d_{\alpha} +
t_{dd}\,[d^{\dagger}_{b} \sigma_{z}d_{a}+h.c]
\label{eq:dspinor}
\end{equation}

\begin{equation}
 H_{T}=\sum_{jk\alpha}\Psi_{jk}^{\dagger} T_{j\alpha} d_{\alpha} + h.c.,
\quad \quad  d_{\alpha}=\left(\begin{array}{c} d_{_\alpha\uparrow}\\
d_{\alpha\downarrow}^{\dagger}
\end{array} \right),
\label{eq:dspinor}
\end{equation}

\noindent 
with $ T_{j\alpha} =t_{j\alpha} \sigma_{z}e^{i\sigma_{z}\varphi_j/2}$ and $t_{j\alpha} $ is
the tunnelling amplitude between the lead $j$ and dot $\alpha$.

The vector connecting the (point) junctions $a-S_0$ and $S_0-b$ is denoted as $\bf R$, and $k_F$ is the Fermi vector in $S_0$. The procedure to obtain the Andreev bound states and the current-phase relationships by writing an effective action for the two dots is found in Ref. [\onlinecite{benjamin}]. One expresses the partition function as 

\begin{equation}
Z = \int \mathcal{D} \left[ \bar{\psi},\psi , \bar{d}, d\right] e^{- S \left[ \bar{\psi},\psi , \bar{d}, d \right]} ,
\end{equation} 

\noindent 
e.g. as a functional integral over Grassmann fields for
the electronic degrees of freedom ($\Psi, \bar\Psi, d, \bar d$). The
Euclidean action reads:

\begin{equation}
S_A=S_{D}+\int_{0}^{\beta} \! d \tau
[\sum_{jk}{\bar\Psi_{jk}(\tau)}(\partial_{\tau}+\xi_{k}\sigma_{z}+\Delta_j\sigma_{x})\Psi_{jk}(\tau)+{
H_{T}(\tau)}]
\end{equation}

\noindent $\beta$ is the inverse temperature, and
${H_{T}(\tau)}=\sum_{jk}{\bar\Psi_{jk\alpha}(\tau)}T_{j\alpha}d_{\alpha}(\tau)+h.c.$
while 

\begin{equation}
S_{D}=\int_{0}^{\beta}\! d\tau [\sum_{\alpha}{\bar d_{\alpha}}(\partial_{\tau}+\epsilon_{\alpha}
\sigma_{z})d_{\alpha}+t_{dd}\,(d^{\dagger}_{b}\sigma_{z}d_{a}+h.c)].
\end{equation}

 After integrating out the leads we get $Z=\int\! \mathcal{D} \left[d_{\alpha}\bar d_{\alpha}\right]\;   \; e^{-S_{eff}}$ 
 with
\begin{equation}
S_{eff}=S_{D}-\int_{0}^{\beta}\! d\tau \; d\tau' \;{\sum_{\alpha \delta}\bar d_{\alpha}(\tau)}{\check
\Sigma_{\alpha \delta}(\tau - \tau ')} d_{\delta}(\tau ')
\end{equation}
where

\begin{eqnarray}
\check
\Sigma_{\alpha\delta}(\tau)=\sum_{j=a,b,0} T_{j\alpha}^{\dagger} G_{j,\alpha \delta}(\tau) T_{j\delta}\\
G_{j,\alpha \alpha}(\tau)=\sum_{k}
(\partial_{\tau}+\xi_{k}\sigma_{z}+\Delta_j\sigma_{x})^{-1}
\delta(\tau)\\
G_{0,a b}(\tau)=\sum_{k}e^{i\bf k\bf R}
(\partial_{\tau}+\xi_{k}\sigma_{z}+\Delta_0\sigma_{x})^{-1}
\delta(\tau).
\end{eqnarray}

We perform a Fourier transform on the Matsubara frequencies (with  $\omega_{n}=(2n+1)\pi/\beta$):
$\delta(\tau)=\frac{1}{\beta} \sum_{\omega_n} e^{-i \omega_{n} \tau}$ and
$G(\tau)=\frac{1}{\beta} \sum_{\omega_n} e^{-i \omega_{n} \tau} G(i\omega_n)$, which gives for the Green's function $G_j$ in terminal $S_j$:

\begin{eqnarray}
\nonumber
G_j(i\omega_n)=\int\! d \xi\; \nu(\xi)(-i
\omega_{n}+\xi_{k}\sigma_{z}+\Delta_j\sigma_{x})^{-1}
\\
\simeq \frac{\pi
\nu(0)}{\sqrt{\Delta_j^2-(i\omega_n)^2}}(i\omega_{n}+\Delta_j \sigma_{x})
\end{eqnarray}

\noindent 
and the nonlocal Green's functions connecting the junctions $a,b$ on the distance $R$ in a one-dimensional channel within terminal $S_0$, 

\begin{align} 
\nonumber
G_{ab}(\omega_n)
\simeq e^{-R/\xi(i\omega_n)}\pi
\nu(0)\\
[\frac{i\omega_{n}+\Delta_0 \sigma_{x}}{\sqrt{\Delta_0^2-(i\omega_n)^2}}\cos(k_FR) + \sigma_{z}\sin(k_FR)].
\end{align}

\noindent 
Here $\xi(i\omega_n)=\frac{\xi_0}{\sqrt{\Delta_0^2-(i\omega_n)^2}}$ and $\nu(\xi)=\sum_{k}\delta(\xi-\xi_{k})$ is
approximated by a constant $\nu(0)$, the density of states at the
Fermi level in the normal leads. Let us set the phase $\varphi_0$ to zero, and assume for sake of simplicity all gaps to be equal, $\Delta_j=\Delta$, and the two junctions equivalent, $t_{aa}=t_{0a}=t_{0b}=t_{bb}=t_{sd}$. This yields the self-energy as a matrix in the Nambu-dots four-dimensional space:

\begin{equation}
\check \Sigma_{\alpha \alpha}(i\omega_n)=\frac{
\Gamma}{2\sqrt{\Delta^{2}-(i\omega_n)^{2}}}[i\omega_{n}-\frac{\Delta}{2}(1+ e^{i\varphi_{\alpha}})\sigma_{x}]
\end{equation}

\begin{eqnarray}
\check \Sigma_{ab}(i\omega_n)=e^{-R/\xi(i\omega_n)}&\nonumber\\
\frac{
\Gamma}{4}[\frac{i\omega_{n}+\Delta \sigma_{x}}{\sqrt{\Delta^{2}-(i\omega_n)^{2}}}\cos(k_FR)+ \sigma_{z}\sin(k_FR)]
\end{eqnarray}
with $\Gamma=2\pi \nu(0)
t_{sd}^2$. Introducing $d_{\alpha}(\tau)=\frac{1}{\sqrt{\beta}}\sum_{\omega_{n}} e^{-i\omega_{n} \tau} d_{\alpha}(i\omega_{n})$ and $\bar {\bf d} = (\bar d_a,\bar d_b)$, we finally obtain the effective action

\begin{eqnarray}  \nonumber
S_{eff}=&\sum_{\omega_{n}} \bar{\bf d}(i\omega_{n}){\bf {\check {\bf M}}}(i\omega_n){\bf d}(i\omega_{n}) \\
{\bf {\check {\bf M}}}(i\omega_n) = &(-i \omega_{n}+\epsilon_{\alpha}
\sigma_{z}) {\check {\bf I}}_{dot}-{\check{ \bf \Sigma}_{i\omega_{n}}},
\end{eqnarray}
where  ${\bf {\check {\bf M}}}(i\omega_n)$ is described by a $4$ x $4$ matrix, whose coefficients are given by

\begin{widetext}
\begin{equation}\label{Matrix} 
\begin{aligned}
M_{11}&=i\omega_n(1+\frac{\Gamma}{2\sqrt{\Delta^2-(i\omega_n)^2}})-E_a,\,\,\,\,\,\,
M_{22}=i\omega_n(1+\frac{\Gamma}{2\sqrt{\Delta^2-(i\omega_n)^2}})+E_a,\\
M_{33}&=i\omega_n(1+\frac{\Gamma}{2\sqrt{\Delta^2-(i\omega_n)^2}})-E_b,\,\,\,\,\,\,
M_{44}=i\omega_n(1+\frac{\Gamma}{2\sqrt{\Delta^2-(i\omega_n)^2}})+E_b,\\
M_{12}&=-\frac{\Gamma \Delta}{4\sqrt{\Delta^2-(i\omega_n)^2}}(1+e^{-i\varphi_a}),\,\,\,\,\,\,
M_{13}=\frac{\Gamma}{4}e^{-R/\xi(i\omega_n)}[\frac{i\omega_n}{\sqrt{\Delta^2-(i\omega_n)^2}}\cos(k_FR)+\sin(k_FR)])+t_{dd},\\
M_{14}&={\check {\bf M}}_{23}=-\frac{\Gamma}{4}e^{-R/\xi(i\omega_n)}[\frac{\Delta}{\sqrt{\Delta^2-(i\omega_n)^2}}\cos(k_FR)]),\\
M_{24}&=\frac{\Gamma}{4}e^{-R/\xi(i\omega_n)}[\frac{i\omega_n}{\sqrt{\Delta^2-(i\omega_n)^2}}\cos(k_FR)-\sin(k_FR)])-t_{dd},\,\,\,\,\,\,
M_{34}=-\frac{\Gamma \Delta}{4\sqrt{\Delta^2-(i\omega_n)^2}}(1+e^{-i\varphi_b}),
\end{aligned}
\end{equation}
\end{widetext}

\noindent
${\bf {\check {\bf M}}}$ being an hermitian matrix once $i\omega_n$ is replaced by the real number $z$. Notice the normal and anomalous couplings between dots, featured by the matrix elements $M_{ij}$ with $i=1,2$ and $j=3,4$. The dispersion relation for the ABS is given by the eigenvalues of the effective action, replacing $i\omega_n$ by $z$. 

 After integrating out the
$\{d_{\alpha},{\bar d_{\alpha}}\}$ variables, the partition function is given by

\begin{equation}
Z=\int\! \mathcal{D} \left[d_{\alpha}\bar d_{\alpha}\right]\;   \; e^{-S_{eff}(d_{\alpha},\bar d_{\alpha})}=\prod_{i\omega_n} \det
{\bf {\check {\bf M}}}(\omega_{n}).
\end{equation}

The free energy reads:
\begin{equation}
 F=-\frac{1}{\beta}\sum_{\omega_n}\ln(\det {\bf {\check {\bf M}}}(i\omega_n)).
\end{equation}

The Josephson current in $S_a$ is expressed as:
\begin{align}
\label{Current}
I_{Ja,b}=\frac{2e}{\hbar}\frac{\partial F}{\partial \varphi_{a,b}}
=-\frac{2}{\beta}\frac{\partial}{\partial \varphi_{a,b}} \sum_{\omega_n} \ln
(\det {\bf {\check {\bf M}}}(i\omega_n))
\end{align}

One can further define an {\it intrinsic} inductance matrix $\bf L$ such as the elements of the inverse inductance matrix $\bf \Lambda = \bf L^{-1}$ are given by :

\begin{equation}
\label{Induct}
\Lambda_{aa}=\frac{\partial I_{Ja}}{\partial \varphi_a},\,\,\,
\Lambda_{bb}=\frac{\partial I_{Jb}}{\partial \varphi_b},\,\,\,
\Lambda_{ab}=\frac{\partial I_{Ja}}{\partial \varphi_b},\,\,\,
\Lambda_{ba}=\frac{\partial I_{Jb}}{\partial \varphi_a}.\,\,\,
\end{equation}

\section{III. Analytical solution in the large gap limit}

In most cases, the contribution to the Josephson current of the continuum states ($|\omega|>\Delta$) is small, therefore one can easily infer the current-phase characteristics from the phase derivatives of the ABS energies. This becomes exact in the so-called large gap limit. One can indeed obtain an analytical solution in the limit $|E_{a,b}|, \Gamma, t_{dd} << \Delta$. This amounts to drop in ${\bf {\check {\bf M}}}$ (Equation \ref{Matrix}) the frequencies $i\omega_n$ in the denominators $\sqrt{\Delta^2-(i\omega_n)^2}$, the factor $i\omega_n$ in ${\check {\bf M}}_{13}, {\check {\bf M}}_{24}$ as well as the renormalization factor $1+\frac{\Gamma}{2\Delta}$ in the diagonal elements. Defining 

\begin{equation}
\label{couplings}
t=\frac{\Gamma}{4}e^{-R/\xi}\sin(k_FR)+t_{dd},\;\;\;\;\bar t=-\frac{\Gamma}{4}e^{-R/\xi}\cos(k_FR),
\end{equation}
one obtains:

\begin{widetext}
\begin{align}
\label{Matrix_grandgap} 
{\check {\bf M}} (i\omega_n)=
\begin{pmatrix}
i\omega_n-E_a & -\frac{\Gamma}{4}(1+e^{-i\varphi_a}) & t & \bar t \\
-\frac{\Gamma}{4}(1+e^{i\varphi_a}) & i\omega_n+E_a & \bar t & -t \\
t & \bar t & i\omega_n-E_b & -\frac{\Gamma}{4}(1+e^{-i\varphi_b}) \\
\bar t & -t & -\frac{\Gamma}{4}(1+e^{i\varphi_b}) & i\omega_n+E_b
\end{pmatrix} 
\end{align}
\end{widetext}
and solving the secular equation $Det({\bf {\check {\bf M}}})=0$ yields the phase dispersion of the ABS cooperatively formed on the two dots, ${\cal E}_{n}=\pm \sqrt{z}$ $(n=1,2,3,4)$ with

\begin{widetext}
\begin{equation}\label{ABSpectrum}
\begin{aligned}
z&=\frac{1}{2}\bigl(E_a^2+E_b^2\bigr) + t^2 + \bar t^2 + \frac{\Gamma^2}{8}\Bigl(\cos^2\frac{\varphi_a}{2}+\cos^2\frac{\varphi_b}{2}\Bigr)\\
&\pm \,\Bigg\lbrace\Bigl[E_a^2-E_b^2 + \frac{\Gamma^2}{8}(\cos\varphi_a-\cos\varphi_b)\Bigr]^2+2\Gamma^2\bar t^2 \Bigl(\cos^2\frac{\varphi_a}{2}+\cos^2\frac{\varphi_b}{2}\Bigr) \\
&+ \Gamma^2\Bigl[t^2\sin^2\bigl(\frac{\varphi_a-\varphi_b}{2}\bigr) - \bar t^2\sin^2\bigl(\frac{\varphi_a+\varphi_b}{2}\bigr)\Bigr] + 8t \bar t \Gamma \Bigl(E_a\cos^2\frac{\varphi_b}{2}+E_b\cos^2\frac{\varphi_a}{2}\Bigr) + 4t^2(E_a+E_b)^2+4 \bar t^2(E_a-E_b)^2\Bigg\rbrace^{\frac{1}{2}}
\end{aligned}
\end{equation}
\end{widetext}

The parameter $t$ reflects the interdot couplings in the normal channel, both through $S_0$ and by direct tunneling (respectively first and second terms in Eq. (\ref{couplings}), and the parameter $\bar t$ represents the anomalous channel through $S_0$. The $S_0$ channels have a dependence in R, both oscillating at the Fermi wavevector and exponentially damped over the coherence length $\xi$. Notice that even in the case where $R >> \xi$ such that nonlocal effects (CAR and EC) are negligible, the interdot coupling plays an essential role, making the bijunction different from two junctions in series. This situation may happen for instance with carbon nanotubes when the central superconducting finger is wide enough but weakly perturbs the nanotube. Let us now discuss the main features of the ABS spectrum within the large gap analytical solution, postponing the general discussion to the next Section. 

\subsection{1. The nonresonant regime}
In the case of uncoupled junctions $(S_0S_a)$, $(S_0S_b)$, e.g. for $t=\bar t=0$, the ABS dispersion for each of the junctions is 
\begin{equation}
{\cal E}_{a(b),\pm}=\pm \sqrt{E_{a(b)}^2+\frac{\Gamma^2}{4}\cos^2\frac{\varphi_{a(b)}}{2}}.
\end{equation}
In the nonresonant regime $\Gamma << |E_{a(b)}|$ it yields a sinuso\"idal current-phase relationship
\begin{equation}
{\cal E}_{a,b,\pm}\simeq \pm [E_{a(b)}+\frac{\Gamma^2}{16|E_{a(b)}|}(1+\cos\varphi_{a,b})].
\end{equation}
If $E_b=\pm E_a$, the ABS in junctions $a,b$ are degenerate. Switching on the nonlocal couplings EC and CAR as well as a possible direct interdot coupling $t_{dd}$ hybridizes the two ABS doublets, yielding a set of four ABS $(n=1-4)$ with ${\cal E}_{1,2} < 0$ and ${\cal E}_{3} = -{\cal E}_{1}, {\cal E}_{4}=-{\cal E}_{2}$, coherently delocalized over the two dots. It is illustrative to perform a perturbative expansion in $\Gamma$ and the interdot couplings $t$, $\bar t$ of expression (\ref{ABSpectrum}), which reduces at $T=0$ to the following approximate expression for the total energy of the bijunction (up to an irrelevant constant) :

\begin{equation}
\label{perturbative_current}
\begin{aligned}
E_{BJ}=&-E_0[\cos\varphi_a+\cos\varphi_b]-E'_0[\cos2\varphi_a+\cos2\varphi_b]\\
&-E_Q\cos(\varphi_a+\varphi_b)-E_{PC}\cos(\varphi_a-\varphi_b).
\end{aligned}
\end{equation}
The first term reflects the "local" tunnel terms of single junctions $a,b$ ($E_0>0$). The second term is the next harmonic, featuring two pairs passing through $a$, or through $b$ ($E'_0<0$). The third and the fourth terms respectively describe quartet tunneling (from $S_0$ towards $S_a,S_b$) and pair cotunneling from $S_a$ to $S_b$. The quartet term is a novel contribution that does not appear in Josephson networks. Expression (\ref{perturbative_current}) yields the inverse inductance $\Lambda_{ab}$

\begin{equation}
\label{perturbative_nlind}
\Lambda_{ab}=\frac{2e}{\hbar}[E_Q\cos(\varphi_a+\varphi_b)-E_{PC}\cos(\varphi_a-\varphi_b)].
\end{equation}

\noindent
On the lines $\varphi_a=\pm \varphi_b=\varphi$ in the ($\varphi_a,\varphi_b$) plane, $\Lambda_{ab}$ oscillates with period $\pi$ with one of the phases (say $\varphi_b$). One obtains  
\begin{equation}
E_Q \approx -\frac{\Gamma^2\bar t^2}{64E_0^4}\;,\;\;\;  E_{PC} \approx \frac{\Gamma^2 t^2}{64E_0^4}
\end{equation}
(assuming $E_a = E_b = E_0$) . One sees that $E_{PC}$ is positive, just as an effective Josephson junction connecting $S_a$ and $S_b$, but on the contrary $E_Q$ is {\it negative}. This means that in terms of quartet tunneling, which depends on the phase combination $\varphi_a+\varphi_b$,  the weakly transparent bijunction is a $\pi$ junction, which here means that the lowest energy is obtained for $\varphi_a+\varphi_b=\pi$.  

This minus sign was discovered  in Ref. \onlinecite{Jonckheere} for the biased bijunction, close to equilibrium, and it comes from the antisymmetry of the Cooper pair wavefunction. Indeed, the quartet mechanism consists in forming two entangled singlet pairs in the dots $a,b$ by a double CAR process. The result of this process is the production of two identical split pairs. Fermion exchange and recombination of these two split pairs into one pair in $S_a$ and one pair in $S_b$ introduces a minus sign. These current components can be probed by applying small voltages $V_{a,b}$ to reservoirs $S_a$, $S_b$ ($V_0=0$) \cite{Freyn,Jonckheere}. Then the phases become time-dependent, $\varphi_a=\varphi_{0a}+\frac{2e}{\hbar}V_a t$ and $\varphi_b=\varphi_{0b}+\frac{2e}{\hbar}V_b t$. In the adiabatic approximation, those time-dependent phases are simply substituted into Equation \ref{perturbative_current}. With $V_{a}=-V_{b}= V$, one obtains the $\pi$-shifted d.c. quartet current $I_Q=-I_{Q0}\sin(\varphi_{a0}+\varphi_{b0})$, which is {\it time-independent}. If one instead fixes $V_{a}=V_{b}=V$, one obtains the coherent pair transfer term $I_{PC}=I_{PC0}\sin(\varphi_{a0}-\varphi_{b0})$, resembling a standard d.c. Josephson term.

Notice that in a strongly nonresonant regime, $t,\bar t,\Gamma << E_{a,b}$, the ABS dispersion becomes independent on the relative signs of $E_a$ and $E_b$. This means that, contrarily to the hybrid splitter $(N_aD_aS_0D_bN_b)$, tuning the levels to $E_a = \pm E_b$ does not help filtering any or the other of EC and CAR processes. This is due to the Andreev reflection which mixes electrons at energy $E$ and holes at energy $-E$. 

\begin{figure}[tb]
\centerline{\includegraphics[width=0.8\columnwidth]{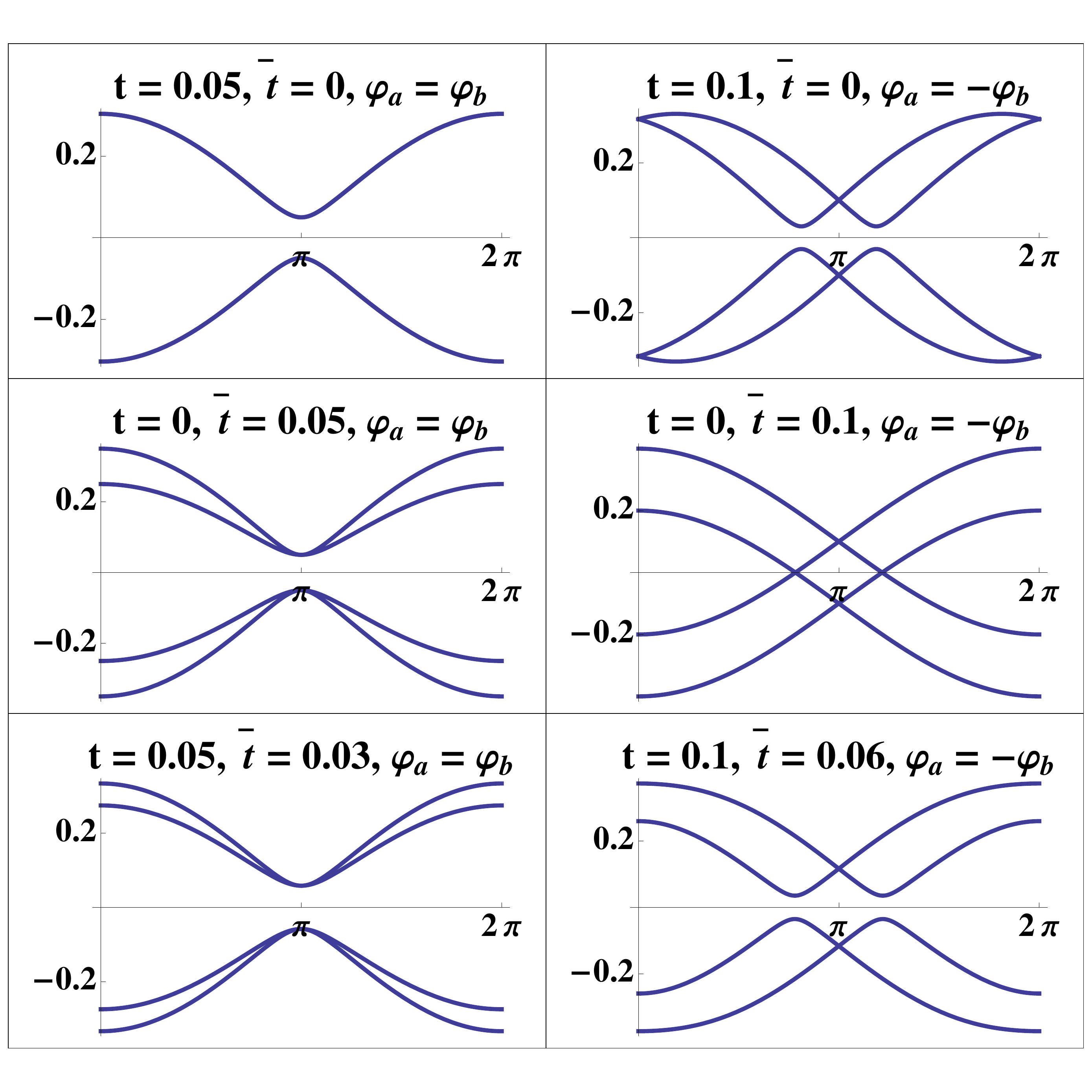}}
\caption{Andreev bound state dispersion $E(\varphi)$  for the bijunction in the resonant case $E_a=E_b=0$, in the large gap approximation, showing the lifting of the degeneracy by the interdot couplings, either at $\varphi=\pi$ along the line $\varphi_a=\varphi_b=\varphi$, or at $\varphi \neq \pi$ on the line $\varphi_a=-\varphi_b=\varphi$. $\Gamma=0.6 \Delta$.
\label{fig:ABSr}
}
\end{figure}

\begin{figure}[tb]
\centerline{\includegraphics[width=0.8\columnwidth]{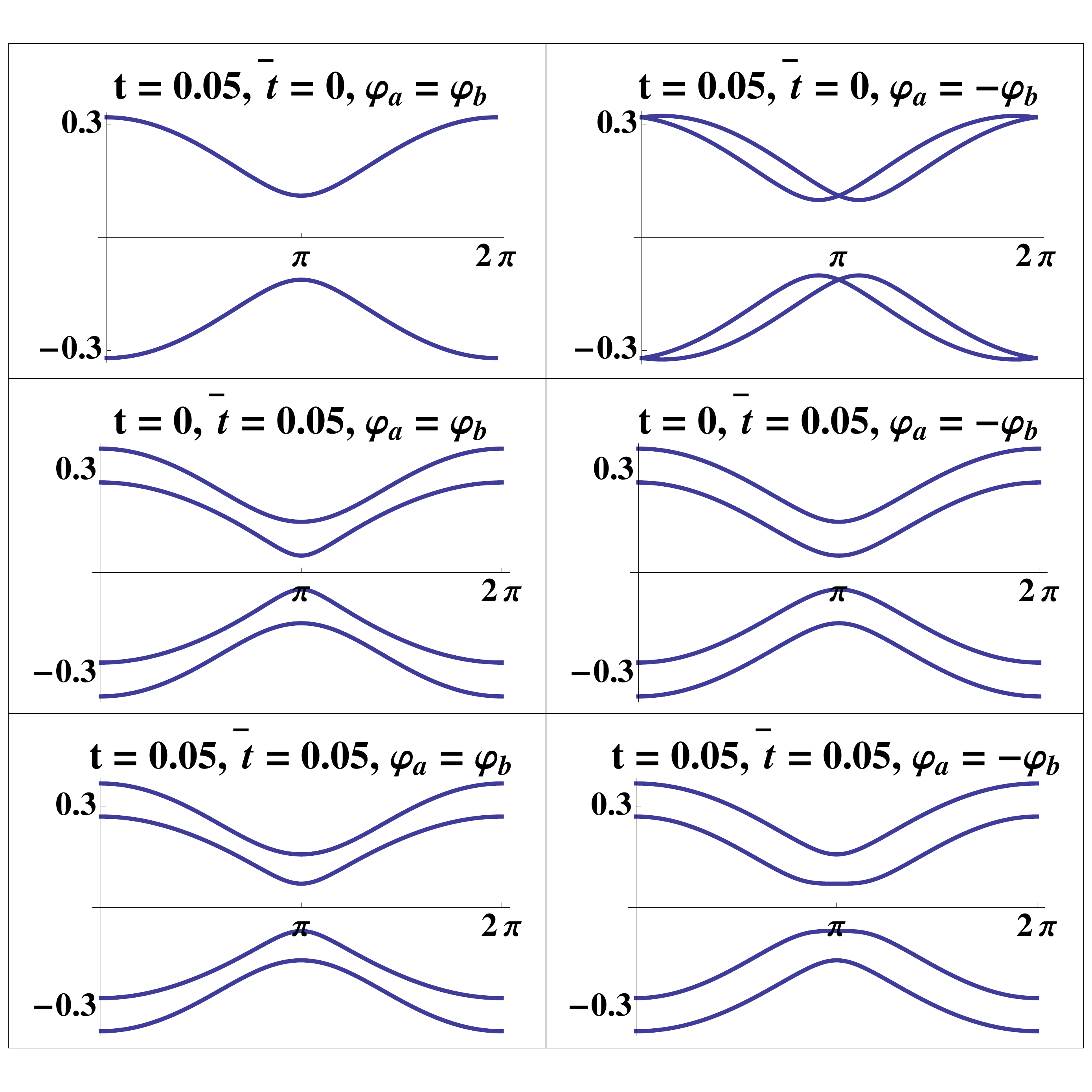}}
\caption{Same parameters as in Figure~\ref{fig:ABSr} but in a slightly nonresonant case, $E_a = -E_b = 0.1\Delta$.
\label{fig:ABSnr}
}
\end{figure}

\subsection{2. The resonant regime}
Let us now turn to the resonant case, $E_a=E_b=0$. Then the ABS dispersion in each junction $a(b)$ alone crosses zero energy at $\varphi_{a(b)}=\pi$. The resulting four-fold degeneracy is lifted by the interdot coupling, in a nonperturbative way. Let us focus on the diagonal directions  in the phase plane.  First, if $\varphi_a=\varphi_b=\varphi$, one finds (one defines $\tilde t=\sqrt{t^2+\bar t^2})$

\begin{equation}
\label{resonant+}
E=\pm\sqrt{\tilde t^2+\big(\frac{\Gamma^2}{4}\pm\Gamma\bar t\big)\cos^2\frac{\varphi}{2}}
\end{equation}

\noindent
showing a structure similar to that of a single dot junction, where $\tilde t$ plays the role of an effective level energy and with an effective coupling $\Gamma\sqrt{1\pm\frac{4\bar t}{\Gamma}}$ if $4\bar t < \Gamma$ which is satisfied from equation (\ref{Matrix}). In the case of no direct interdot coupling, $\tilde t=\frac{\Gamma}{4}e^{-R/\xi_0}$ does not depend on the geometrical phase $\beta_R=k_FR$, contrarily to the couplings $t$ and $\bar t$ separately. Equation \ref{resonant+} can be interpreted in terms of "molecular states" formed on the double dot, due to the interdot couplings $t$ and $\bar t$ (direct and through CAR and EC)  with a degeneracy lifted by the local couplings to the superconductors, represented by $\Gamma$ (Figure \ref{fig:ABSr}, left panels). The scale of the splitting is given by $\tilde t$. 

On the other hand, in the case $\varphi_a=-\varphi_b=\varphi$, one obtains:

\begin{equation}
\label{resonant-}
E=\pm\sqrt{\tilde t^2+\frac{\Gamma^2}{4}\cos^2\frac{\varphi}{2}\pm\Gamma|\cos\frac{\varphi}{2}|\big(\bar t^2+t^2\sin^2\frac{\varphi}{2}\big)^{1/2}}
\end{equation}

In the peculiar case $t=0$, which can be achieved if $t_{dd}=0$ and $\beta_R=n\pi$, the dots are coupled only in the electron-hole channel, and the solution presents two twofold degenerate crossing points $E=0$, at $\varphi=\pi \pm 2\arcsin\big(\frac{2\bar t}{\Gamma}\big)$. Coupling in the electron-electron channel by the parameter $t$ lifts this degeneracy, leaving a two-gap structure (Figure \ref{fig:ABSr}, right panels). This kind of degeneracy lifting is qualitatively different from that encountered along the other diagonal $\varphi_a=\varphi_b$, where the $E=0$ crossing instead occurs at ($\pi,\pi$). Indeed the scale of the phase splitting of the crossing points is given by $\tilde t/\Gamma$. Yet the energy splitting at those crossing points is of the order of $2\frac{t}{\Gamma}\tilde t$, thus these minigaps are much smaller than the one formed at $\varphi=\pi$ in the case $\varphi_a=\varphi_b$. To complete this picture, a case close to resonance is represented in Figure \ref{fig:ABSnr}.

Several remarks must be made in the resonant regime. First, it is no more possible to distinguish between quartet and pair cotunneling processes. Just as in a single transparent SNS junction couples two superconductors by a strongly nonperturbative proximity effect in the N region, the bijunction ensures a coupling between three superconductors by proximity effect in the double dot. Second, due to lifting of the four-fold degeneracy, the sharp qualitative change between the individual ABS and the full bijunction structure holds at $T=0$ for any, whatever weak, interdot coupling, including the case of a wide ($R >> \xi$) central superconductor. 

\section{IV. General discussion}

\subsection{1. Current-phase relationships and the nonlocal inductance.}

Let us now discuss the numerical results from Equations \ref{Current}, \ref{Induct}, without the large gap approximation. The current-phase relationships $I_a(\varphi_a,\varphi_b), I_b(\varphi_a,\varphi_b)$ and the inverse inductance matrix $\Lambda_{ij}$ can be exactly obtained, both at zero and at finite temperature. Compared to uncoupled junctions $(SS_a)$, $(SS_b)$, the cuts of the $I(\varphi_a,\varphi_b)$ along the directions $\varphi_a=\varphi_b$ (resp. $\varphi_a=-\varphi_b$) are dominated by the quartet (resp. pair cotunneling) contributions and their harmonics.  

\begin{figure}[tb]
\centerline{\includegraphics[width=0.8\columnwidth]{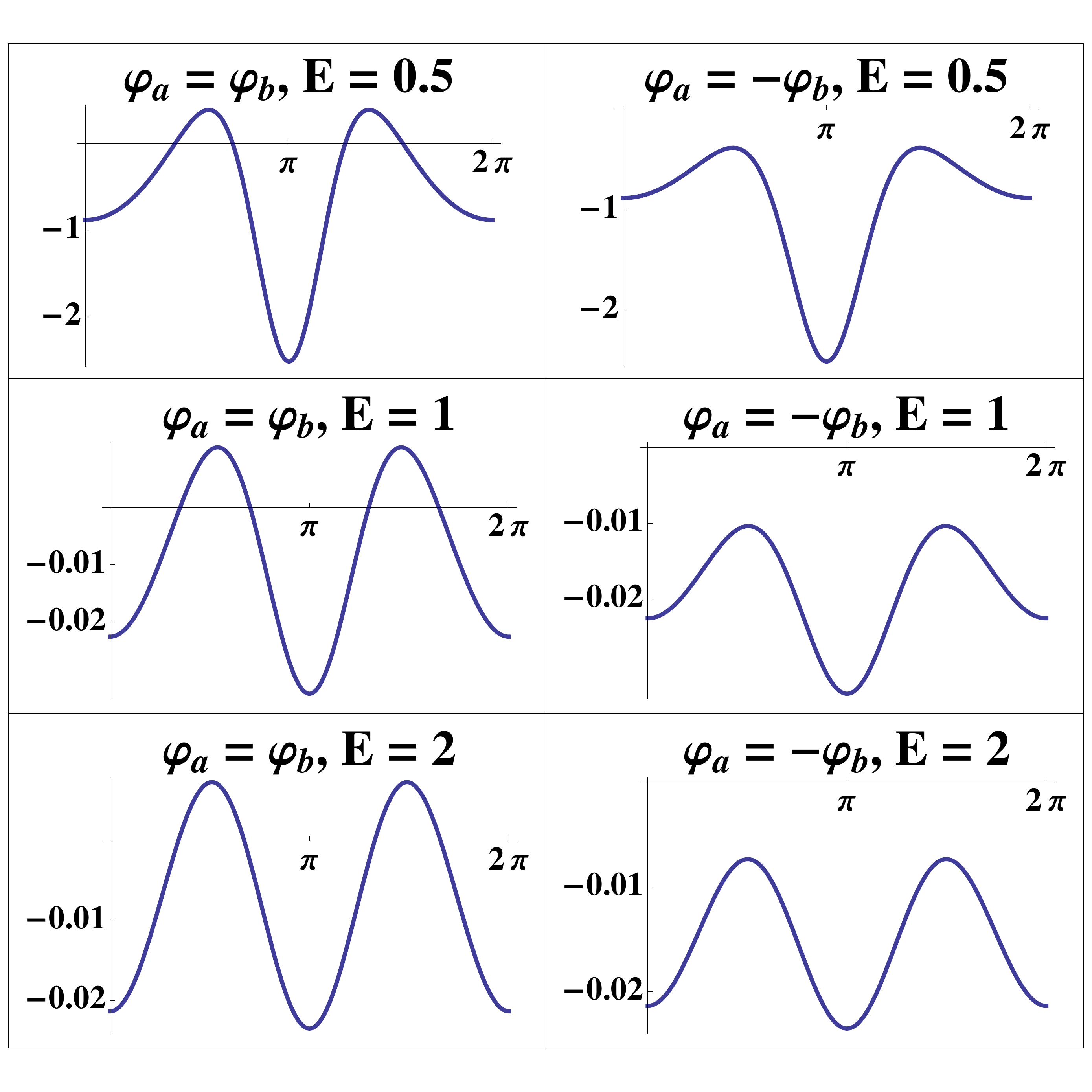}}
\caption{Exact solution : nonlocal inverse inductance $\Lambda_{ab}(\varphi_b)$ (x $10^3$) in the nonresonant regime, for $\Gamma=\Delta$, $T=0.05\Delta$, $R=\xi$, $\beta_R=\frac{\pi}{7}$ and $E_a=\pm E_b=E$.
\label{fig:Indnlnonres}
}
\end{figure}

The results depend on the values of the dot couplings $t,\bar t$, e.g. of the  phase $\beta_R$, when there is no direct coupling $t_{dd}$. For instance, fixing $\beta_R=\frac{\pi}{7}$, both CAR and EC processes contribute. On the other hand, fixing $\beta_R=\frac{\pi}{2}$, EC dominates, and fixing $\beta_R=\pi$, CAR dominates. 

In the nonresonant regime, Figure \ref{fig:Indnlnonres} shows the exact result for the inverse nonlocal inductance, approaching the $-\cos(2\varphi)$ regime for large dot energies. Comparing to the perturbative expression equation \ref{perturbative_nlind}, it is clear from this figure that $E_{PC}>0$ but $E_Q<0$, generalizing the analytical large gap result of Section III.

The resonant regime displays a strong anharmonicity. Figure \ref{fig:Crt} shows $I_a(\varphi_a=\varphi,\varphi_b=\varphi)$, $I_a(\varphi_a=\varphi,\varphi_b=-\varphi)$ and $\Lambda_{ab}(\varphi_a=\varphi,\varphi_b=\varphi)$, $\Lambda_{ab}(\varphi_a=\varphi,\varphi_b=-\varphi)$. The effect of the interdot coupling is apparent in the $I_a(\varphi_b)$ plots for $\varphi_b=\pm\varphi_a$. One takes as a reference the current $I_{a0}(\varphi_b)=\pm I_{a0}(\varphi_a)$ in absence of interdot coupling and nonlocal effects. For $\varphi_a=\varphi_b$ the nonlocal processes opening a gap at phase $\pi$ (Figure \ref{fig:ABSr}) smoothen the current jump, and are dominated by a quartet $\pi$-component. For $\varphi_a=-\varphi_b$ the splitting of the crossing points give rise to a double jump, showing the nonperturbative nature of CAR and EC couplings.

Similarly, the inductance features shown in Figure \ref{fig:Crt} can be understood qualitatively from the "large gap" ABS spectra calculated in Section III (Figure \ref{fig:ABSr}). The negative peak in $\Lambda_{ab}(\varphi_b)$ along the line $\varphi_a=\varphi_b$ comes from the splitting of the individual ABS by the interdot coupling (Figure \ref{fig:ABSr}, left panels). 
It has a modified Lorentzian shape, and at zero temperature and for $\tilde t << \Gamma$ its width scales as $\tilde t/\Gamma$ and its height scales as $\Gamma^2/\tilde t$. On the other hand, along the line $\varphi_a=-\varphi_b$, the two positive and very sharp symmetric peaks originate from the splitting of the ABS crossing along the phase axis (Figure \ref{fig:ABSr}, right panels). 
The splitting scales as $\tilde t/\Gamma$. The divergence of the nonlocal inductance when the interdot coupling goes to zero is an effect of a degeneracy lifting. It disappears at nonzero temperature, which smoothens all the above structures when $\beta \tilde t < 1$. Once more, notice that the results for $E_a = E_b$ and $E_a = -E_b$ are not very different. In particular, taking $E_a = E_b$ does not filter out the CAR processes, just as taking $E_a = -E_b$ does not filter out the EC processes.

\begin{figure}[tb]
\centerline{\includegraphics[width=1.\columnwidth]{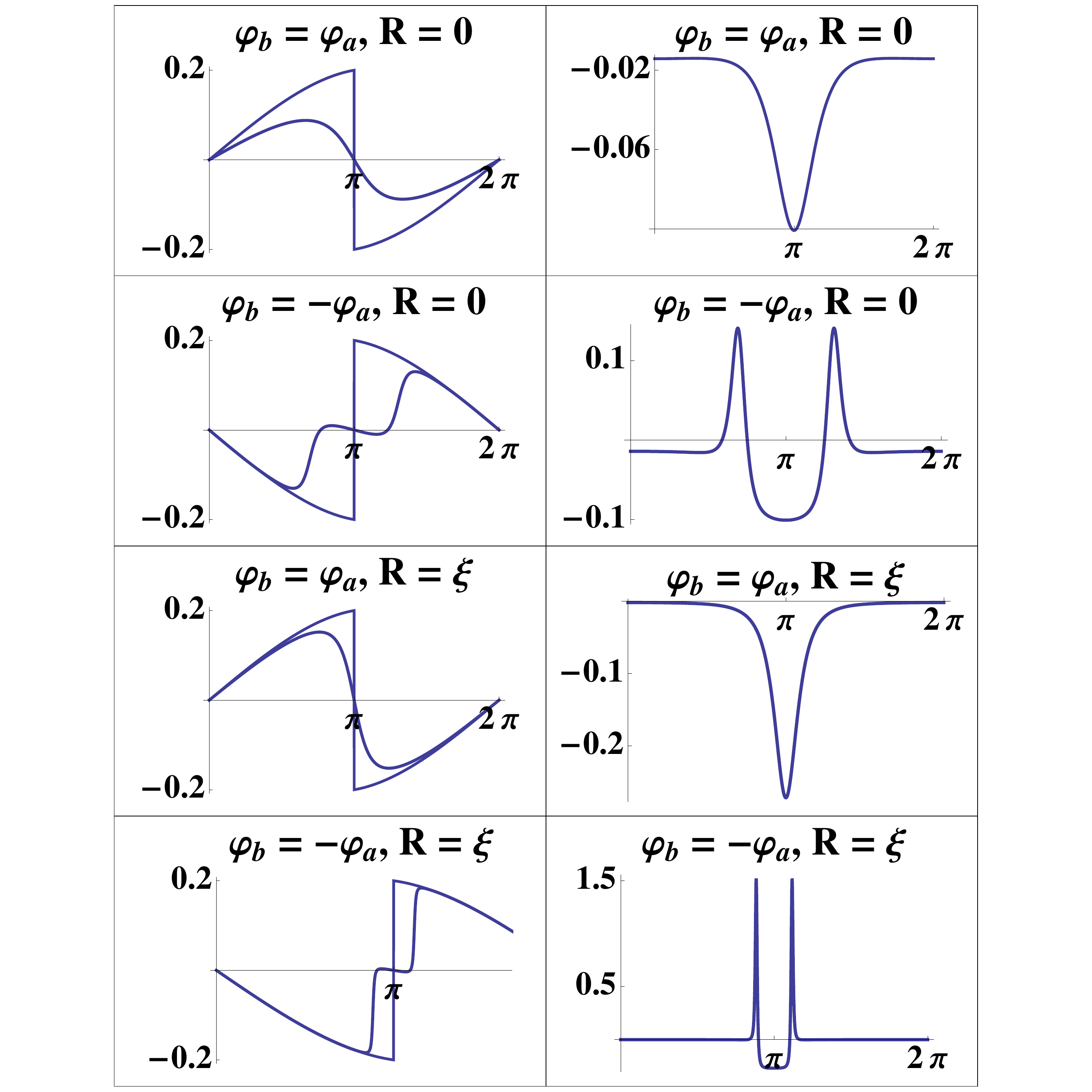}}
\caption{Current $I_a$ as a function of $\varphi_b$ (left panels) and nonlocal inverse inductance $\Lambda_{ab}=\frac{\partial I_{Ja}}{\partial \varphi_b}$ (right panels) in the resonant dot case, for $\varphi_b=\pm\varphi_a$, for strong ($R=0$) and intermediate ($R=\xi$) interdot coupling through $S_0$. The local resonant current (with a sharp drop at $\pi$) is plotted as a reference. Temperature is zero, $E_a=E_b=0$ and the geometrical phase is $\beta_R=\frac{\pi}{7}$. Notice the sharpening of the structures in $\Lambda_{ab}$ as the interdot coupling weakens.
\label{fig:Crt}
}
\end{figure}




\subsection{2. Effect of the circuit inductance}
In a circuit where the bijunction is closed by two adjacent loops (Figure \ref{TwoLoop}), the geometrical inductance matrix of the circuit should be taken into account, ${\bf L}_0={L_{0aa},L_{0bb},L_{0ab}=L_{0ba}=M_0}$. In particular, the mutual inductance $M_0$ couples the pair currents in junctions $a$ and $b$, and it could interfere with the detection of the quartet and  pair cotunneling processes. 

\begin{figure}[tb]
\centerline{\includegraphics[width=1.\columnwidth]{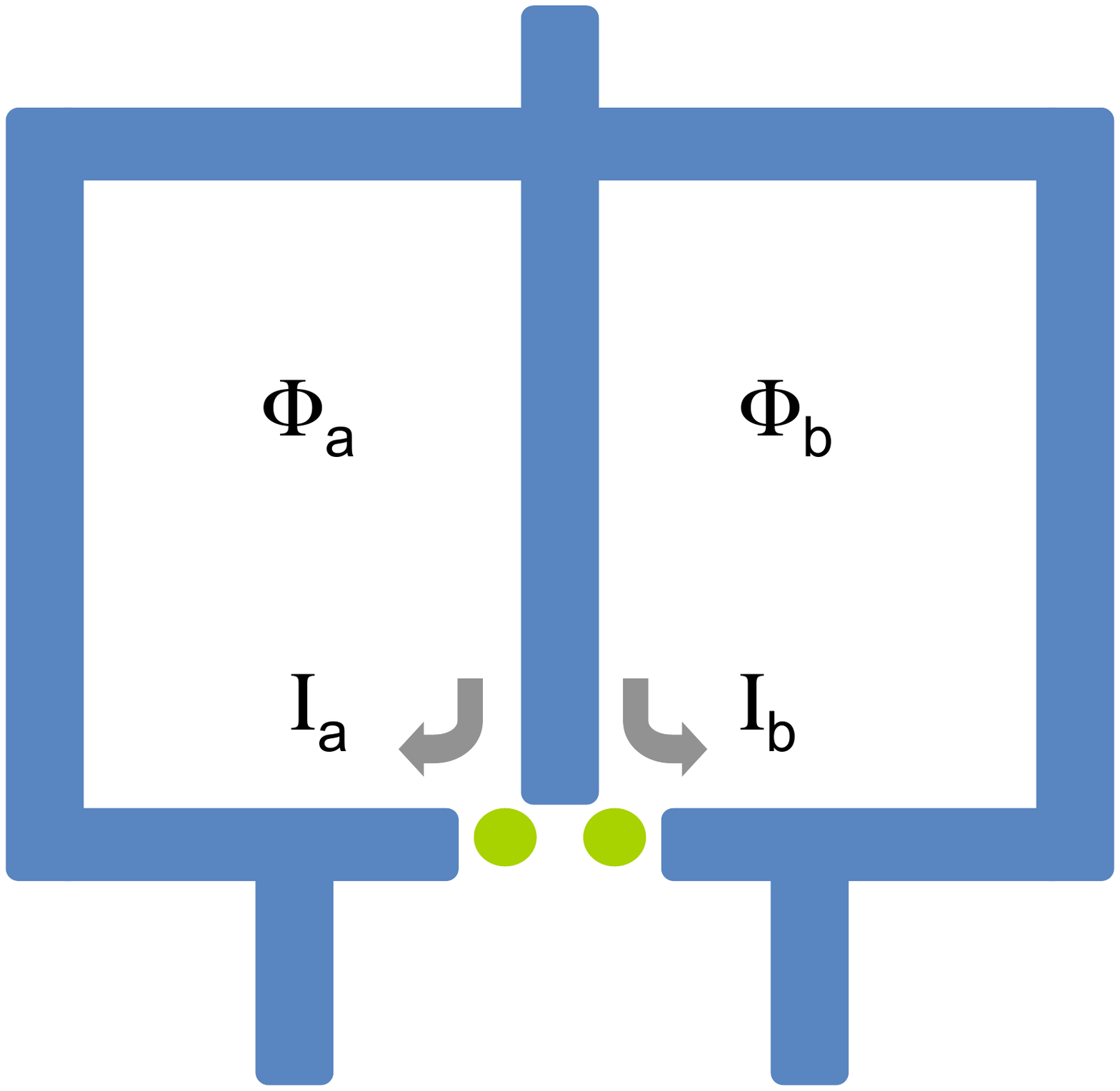}}
\caption{Scheme of a double dot bijunction inserted into a two-loop and tri-terminal circuit. 
\label{TwoLoop}
}
\end{figure}

Let us consider the double loop circuit pictured in Figure (\ref{TwoLoop}). The convention of currents flowing from the central superconductor to the side ones amounts to change the sign of $\varphi_b$ and $I_b$, therefore the phase differences $\varphi_a$ and $\varphi_b$ are related to the external fluxes $\Phi_{ea}$ and $\Phi_{eb}$ in loops $(a,b)$ by ($\Phi_0=\frac{hc}{2e}$) :

\begin{equation}
\begin{aligned}
\varphi_a&=\frac{2\pi}{\Phi_0} (\Phi_{ea}+L_{0aa}I_a-M_0I_b)\\
\varphi_b&=-\frac{2\pi}{\Phi_0} (\Phi_{eb}-L_{0bb}I_b+M_0I_a)
\end{aligned}
\end{equation}

\begin{figure}[tb]
\centerline{\includegraphics[width=0.9\columnwidth]{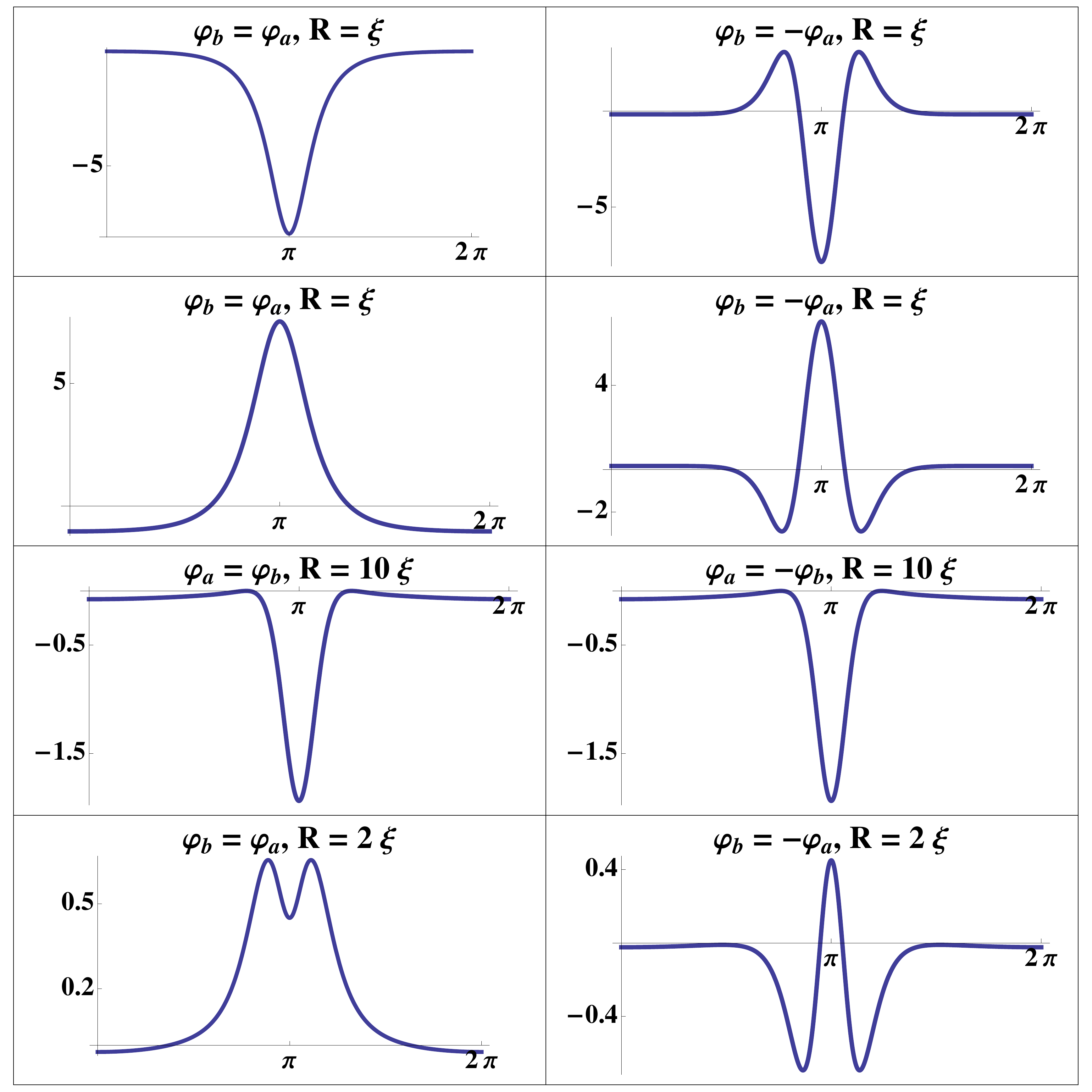}}
\caption{Effect of self and mutual inductances on the nonlocal inverse inductance (scale x$10^2$) for $E_a=E_b=0, T=0.02 \Delta$. Top line panels: reference curves with nonlocal couplings and no inductance, plotting $\Lambda_{ab}(\varphi_b)$; Second line panels: with nonlocal couplings and self, $L_{0aa}=L_{0bb}=0.2$, plotting $\Lambda_{eab}(\Phi_b)$; Third line panels: with self $L_{0aa}=L_{0bb}=0.2$ and mutual $M_0=-0.1$, without nonlocal couplings, plotting $\Lambda_{eab}(\Phi_b)$; Fourth line panels: with self $L_{0aa}=L_{0bb}=0.2$ and mutual $M_0=-0.06$, with nonlocal couplings, plotting $\Lambda_{eab}(\Phi_b)$.
\label{fig:mutuelle}
}
\end{figure}

\noindent 
We define the full inverse nonlocal inductance as $\Lambda_{eij}=\frac{2e}{\hbar}\frac{\partial I_i}{\partial \Phi_{ej}}$, $(i,j=a,b)$. 
Figure (\ref{fig:mutuelle}) compares this quantity to the one due only to nonlocal couplings, and shows it for several cases. With the self $L_{0aa},L_{0bb}$ and with nonlocal coupling, the patterns $\Lambda_{eab}(\Phi_b)$ are qualitatively similar to the patterns $\Lambda_{ab}(\varphi_b)$, but inverted owing to the phase and flux sign convention. With the mutual inductance $M_0$ in addition, but without nonlocal coupling, the pattern is inverted compared to the previous one. This is due to the fact that the mutual inductance is {\it negative}, e.g. it tends to make the currents flowing in loops $a,b$ cancel in the common branch, while the quartet process favours the {\it same} sign for the currents. Finally, with both nonlocal coupling and mutual inductance, the former is distincly visible, with a dip in the left panel. The marked difference between the two lowest panels of Figure \ref{fig:mutuelle} shows that for a realistic circuit the nonlocal processes can be distinguished from the geometric inductances. An alternative to fiter out the purely geometric effects is to modulate one or the other of the couplings and operate a synchronous detection.

\section{conclusion}

\begin{figure}[tb]
\centerline{\includegraphics[width=0.9\columnwidth]{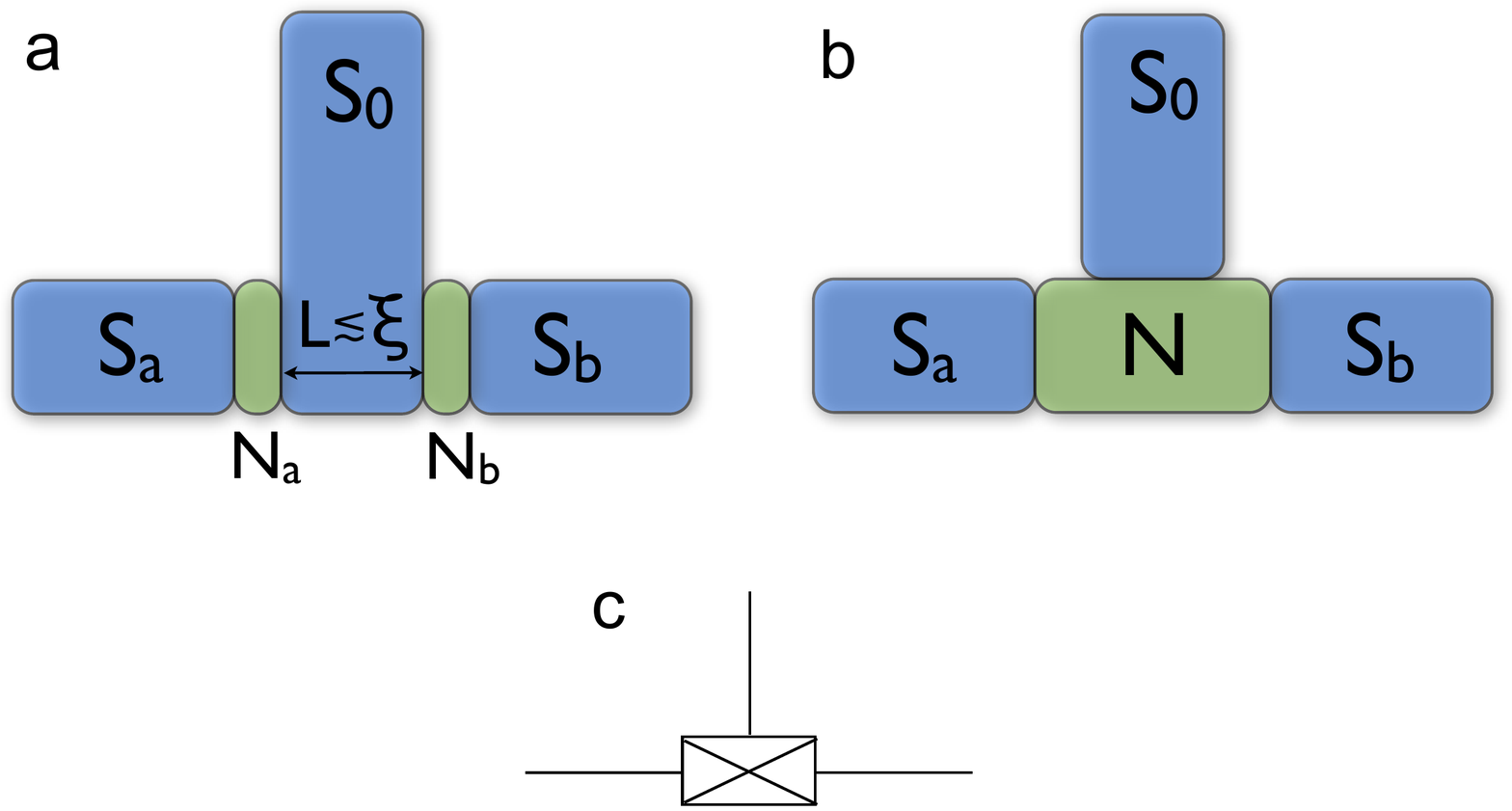}}
\caption{a) Scheme of a bijunction made of three superconductors coupling $S_0$ to $S_a$ and to $S_b$ through normal regions $N_a$, $N_b$. Coupling between $S_a$ and $S_b$ is mediated by nonlocal processes through $S_0$. b) Bijunction with $S_0$, $S_a$, $S_b$ all mutually coupled through a normal region $N$. c) Pictorial circuit element scheme for a bijunction. 
\label{Bijunctions}
}
\end{figure}

We have calculated the (two current)-(two phase) characteristics of a double dot bijunction, unveiling the anharmonicities occurring in the resonant and degenerate dot level case. The approximate and exact calculations presented in this work enlighten the nature of the proximity effect induced by three superconductors on a double dot forming a Josephson bijunction. We have emphasized the role of the interdot coupling even when the central superconductor is too wide to mediate nonlocal effects. Even a weak coupling between the two junctions, mediated by the central superconductor or by direct interdot tunneling, has strong effects, inducing a measurable nonlocal inductance of purely microscopic origin. In case of a two-loop circuit, it has the opposite sign compared to a geometrical mutual inductance. Alternatively, the current-phase structure can be directly investigated through recently introduced spectroscopy techniques. The Andreev bound state structure is also a necessary basis for understanding the more complicated nonequilibrium behaviour, as investigated in Ref. \onlinecite{Jonckheere}. One has to keep in mind nevertheless that the usual adiabatic approximation fails unless the voltages are small enough, and at any voltage in the resonant regime. The phenomenology revealed in a double dot bijunction can be generalized to bijunctions formed with normal metal regions, that can be disconnected (Figure \ref{Bijunctions}a) or connected (Figure \ref{Bijunctions}b).

\begin{acknowledgments}
We acknowledge the support of the French National Research Agency, through the project ANR-NanoQuartets (ANR-12-BS1000701). This work has been carried out in the framework of the Labex Archim\`{e}de (ANR-11-LABX-0033) and of the A*MIDEX project (ANR-11-IDEX-0001-02), funded by the ``Investissements d'Avenir'' 
French Government program managed by the French National Research Agency (ANR). We are grateful to T. Kontos for useful discussions.
\end{acknowledgments}

\end{document}